\documentclass[conference]{IEEEtran}
\usepackage{amsmath}
\usepackage{txfonts}
\usepackage{graphicx}
\usepackage{cite}
\usepackage{epstopdf}

\usepackage{mathabx} 

\ifCLASSINFOpdf
\else
\fi
\usepackage{hyperref} 
\hypersetup{colorlinks,citecolor=black,filecolor=black,linkcolor=black,urlcolor=blue}

\begin{document}
%
\title{DEUS Full Observable $\Lambda$CDM Universe Simulation: the numerical challenge}
\author{\IEEEauthorblockN{Jean-Michel Alimi, Vincent Bouillot\\
Yann Rasera, Vincent Reverdy,\\ 
Pier-Stefano Corasaniti \& Ir\`ene Balm\`es }
\IEEEauthorblockA{LUTh, CNRS, Observatoire de Paris, Univ. Paris Diderot ;\\ 
5 Place Jules Janssen, 92190 Meudon, France.\\
http://www.deus-consortium.org}
\and
\IEEEauthorblockN{St\'ephane Requena}
\IEEEauthorblockA{GENCI\\
12, rue de l'Eglise, \\
75015 Paris, France.\\
http://www.genci.fr}
\and
\IEEEauthorblockN{Xavier Delaruelle\\
\& Jean-Noel Richet}
\IEEEauthorblockA{TGCC, CEA DAM, DIF\\ 
91297 Arpajon, France.\\
http://www-hpc.cea.fr}}


%


\maketitle

\begin{abstract}
We have performed the first-ever numerical N-body simulation of the full observable universe (DEUS "Dark Energy Universe Simulation" FUR "Full Universe Run"). This has evolved 550 billion particles on an Adaptive Mesh Refinement grid with more than two trillion computing points along the entire evolutionary history of the universe and across 6 order of magnitudes length scales, from the size of the Milky Way to that of the whole observable universe. To date, this is the largest and most advanced cosmological simulation ever run. It provides unique information on the formation and evolution of the largest structure in the universe and an exceptional support to future observational programs dedicated to mapping the distribution of matter and galaxies in the universe. The simulation has run on 4752 (of 5040) thin nodes of BULLÕ supercomputer CURIE, using more than 300 TB of memory for 10 million hours of computing time. About 50 PBytes of data were generated throughout the run. Using an advanced and innovative reduction workflow the amount of useful stored data has been reduced to 500 TBytes.
\end{abstract}


%
\IEEEpeerreviewmaketitle

\section{Introduction}
Over the past decades cosmological observations have provided us with a clearer picture of the universe \cite{Perlmutter1999,Riess1998,York2000,Spergel2003,Komatsu2009}. 
On the one hand these measurements have confirmed the pillars of the Standard Hot Big-Bang scenario \cite{Peebles1993}, on the other hand they showed the existence of a class of phenomena whose study has opened a window on a completely new and unexplored territory in physics. The atoms which makes planets, stars, the diffuse gas between galaxies
and ourselves account only for a small fraction of the total content of the universe. Quite astonishingly $95\%$ of the cosmos is in the form of two invisible components: $22\%$ is in Cold Dark Matter (CDM) particles, which are primarily responsible for the formation of the visible structures in the universe \cite{Peebles1980}; $73\%$ is in a unknown exotic form, dubbed ``dark energy'' (DE), which is responsible for the present phase of cosmic accelerated expansion \cite{Copeland2006}. 

Super-symmetric particles and extensions of the Standard Model of particles physics can provide physically motivated candidates to the role of Dark Matter (DM). In contrast, there is little clue on Dark Energy (DE). The presence of a cosmological constant term $\Lambda$ in Einstein's equations of General Relativity \cite{Weinberg1972} can account for the observed phenomenon and the so-called "concordance'' $\Lambda$CDM scenario has emerged as the minimal model to fit all available cosmological observations. However, the measured value of $\Lambda$ is hardly reconcilable with any physical interpretation so far proposed. As a result, Dark Energy may well be of different origin. Several alternative scenarios have been advanced, but to date a coherent physical understanding of DE is still missing. 

In the lack of a theoretical guidance, cosmologists are turning to observations and in the future several observational projects such as BOSS\cite{BOSS}, DES\cite{DES}, LSST\cite{LSST} and the EUCLID mission\cite{EUCLID} will map with great accuracy the distribution of matter in the universe. These may eventually shed new light on the problem as the clustering of matter may be key to infer the properties of DE.
Indeed the physics responsible for DE may alter the time evolution of the clustering of DM which in turn shapes the formation of the visible structures in the universe.

{\it What imprints does Dark Energy leave on the cosmic structures? And inversely, how can the nature of Dark Energy be inferred from observations of the distribution of matter in the universe?}

These are the fundamental questions that the Dark Energy Universe Simulation\cite{DEUS}
(DEUS) project seeks to answer. 

The galaxy and clusters we observe today are the result of the growth of tiny density fluctuations present in the early universe and which we observe today as temperature fluctuations in the Cosmic Microwave Background (CMB) radiation. The gravitational infall of initial DM density fluctuations has evolved over time to reach a highly non-linear dynamical regime in which DM particles bound into stable objects, the halos. It is inside DM halos that cooling gas falls in to form stars and galaxies and it is the succession of halo mergers that shapes the final distribution of the large scale structures we observe today. Therefore, in order to study the imprint of Dark Energy on the cosmic structure formation one has to follow the gravitational collapse of Dark Matter throughout the history of the universe and across several order of magnitude length scales. This is not realizable using solely analytical methods, which break down as soon as non-linearities develop in the dynamics of DM. Only numerical N-body simulations provide the tool
to follow the entire evolution of Dark Matter particles in an expanding universe dominated by Dark Energy. 

During the past 10 years several groups have pushed to the limits both size and resolution of cosmological N-body simulations. The Millenium Simulation in 2005 has run a $2.2$ billion light-years simulation box with $10$ billion particles \cite{Springel2005}. Since then the performance of cosmological simulations has rapidly increased. The Millenium-XXL simulation has evolved more recently 303 billion particles in a 13 billion light-years box \cite{angulo12}, while the Horizon Run 3 has followed the evolution of 374 billion particles in a 49 billion light-years box.

A simulation of the size of the entire observable universe has been a long dreamed goal to infer cosmic variance limited predictions of the properties of the distribution of cosmic structures.

In this paper, we present the first-ever numerical simulation of the entire observable universe, from the Big Bang to the present day for the concordance $\Lambda$CDM model. This simulation has followed the gravitational infall of 550 billion particles in a 95 billion light-years simulation box (assuming the adimensional Hubble constant to be $h=0.72$), the size of the entire observable universe. The realization of this simulation, 
the first of the three planned runs of the DEUS Full Universe Runs (DEUS FUR) project, has used the totality of the CURIE supercomputer provisioned by the Grand \'Equipement National de Calcul Intensif \cite{GENCI} (GENCI) and operated at the "Tr\`es Grand Centre 
de Calcul" (TGCC) of "Commissariat \`a l'\'Energie Atomique et aux \'Energie Alternatives" \cite{CEA} (CEA). The other two planned runs of Dark Energy models alternative to the $\Lambda$CDM are scheduled for the next few weeks.

\begin{figure}[b]
	\centering
	\includegraphics[scale=0.3]{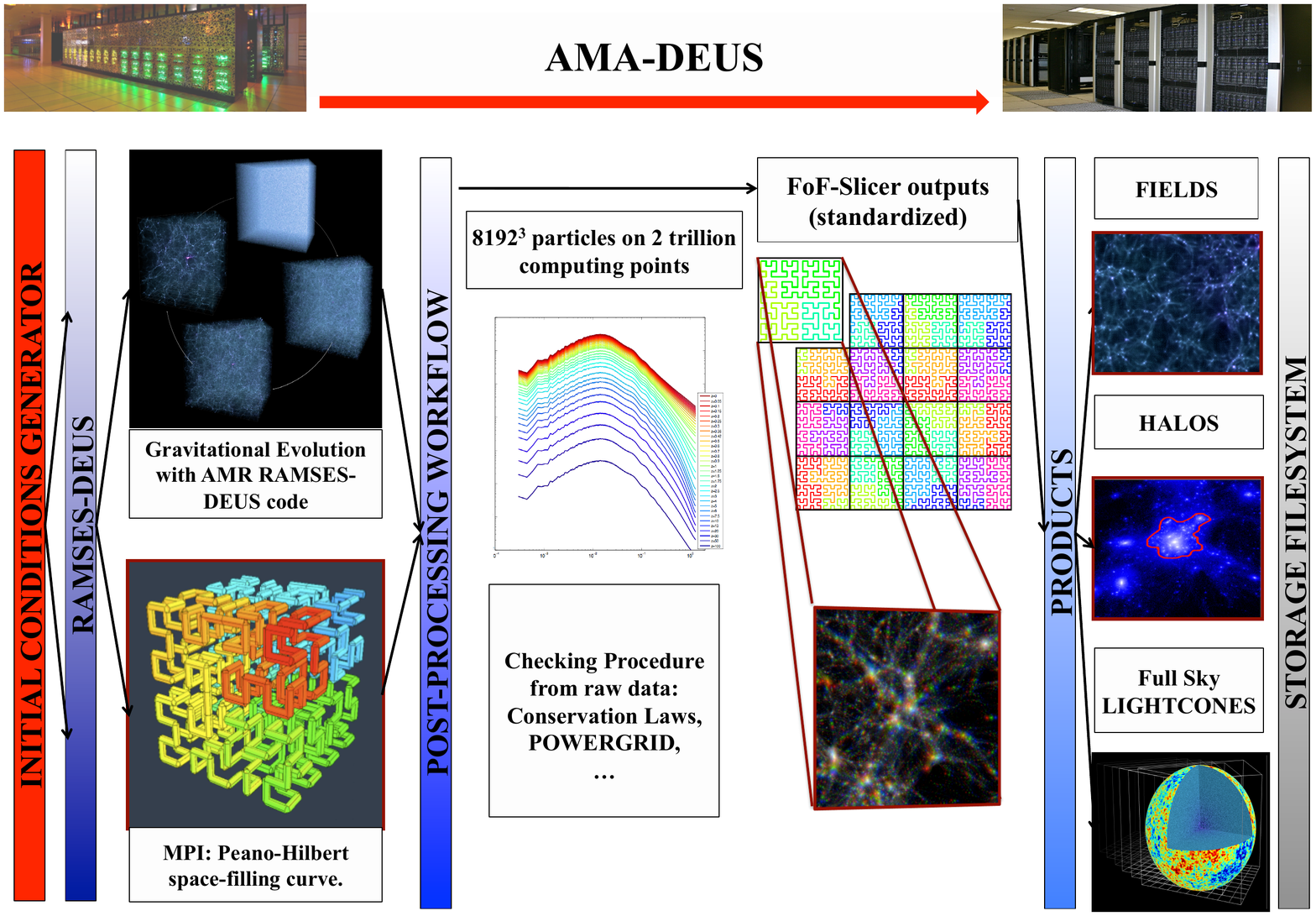}
	\caption{A Multiple purpose Application for Dark Energy Universe Simulation "AMA-DEUS": Four components: (i) Generator of initial condition, (ii) Dynamical solver of gravitational clustering, (iii) Post-processing, (iv) Numerical products and storage. }
	\label{AMADEUS}
\end{figure}

Here, we will describe the main technical challenges that our team has faced to successfully run
the first simulation of the full observable universe. To date this is the most advanced cosmological simulation ever realized. The paper is organized as follows. In Section 2 we briefly review the science implications of N-body simulations in cosmology and compare the DEUS FUR to previous "Grand Challenge'' simulations. In Section 3 we present ``A Multiple purpose Application for Dark Energy Universe Simulation'' (AMA-DEUS) that has been developed specifically for this project (a schematic representation of the application is shown in Figure \ref{AMADEUS}. This application includes the generator of the initial conditions, the numerical algorithms that have been used to solve the gravitational evolution of Dark Matter particles and an innovative reduction workflow to drastically reduce the data stored during the run. Without such a reduction workflow, it would have been impossible to perform the runs. The CURIE supercomputer is briefly presented in Section 4, while in Section 5 we describe the optimization of the numerical codes, the performances that we have been able to obtain and the numerical applications that we have used. Finally, in Section 6 we highlight some preliminary scientific results and present our conclusions.

\section{DEUS FUR Science \& Characteristics}
The goal of the DEUS FUR simulations is to shed light on the imprint of Dark Energy on the cosmic structure formation. To this purpose, numerical N-body simulations of the entire observable universe are run for three different Dark Energy models. These consists of a standard $\Lambda$CDM model and two DE scenarios which are alternative to the $\Lambda$CDM and have characteristics complementary to each other: a late-time slowly evolving dynamical "quintessence'' scalar field and a "phantom'' DE fluid. 
The parameters of these models have been calibrated to the latest observations of the CMB and in agreement with cosmic distance measurements such that the models are statistically indistinguishable from each other. This allows us to test whether future probes of the non-linear clustering of DM may distinguish between these competing models and hence shed new light on the DE problem. A comparative analysis of the DM distribution from the three model simulations will allow us to identify the imprints of DE on the non-linear structure formation. However, disposing of accurate numerical results is a necessary condition for the success of this study. For instance, if we consider the cosmological evolution of the DM density fluctuations spectrum we would like to best resolve the scale of $\sim 100\,{\rm h^{-1}}$ Mpc (1 parsec = 3.26 light-years), where the feature of the Baryon Acoustic Oscillations (BAO) appears. This is the imprint of the acoustic waves propagating in the primordial plasma carrying a signature of the late time cosmic accelerated expansion. 
However, in order to obtain cosmic variance limited predictions the simulation box must have the size of the full observable universe $\sim 21\,{\rm h^{-1} Gpc}$, which in physical units (for $h=0.72$) contains roughly 210 times the BAO scale. Moreover, we would like to have also a sufficient mass resolution such as to resolve the most massive clusters, those with masses above $10^{14}$ M$_\odot$ (solar masses). This requires that halos in the simulations contain at least 100 particles. Furthermore, by simulating the full observable universe, we are guaranteed to detect the rare super-massive clusters in the universe whose exact number is a sensitive indicator of DE. Hence, to achieve the desired mass (\mbox{$>10^{12}$ M$_\odot$}) and spatial ($>40\,{\rm h^{-1} kpc}$) resolution, the DEUS FUR simulations have $\sim 550$ billions particles whose evolution is evaluated along the entire history of the universe using an Adaptive Mesh Refinement grid with more than two trillion computing points.

The impact of these simulations goes beyond the study of the physical effects of DE on the cosmic structure formation. In fact, by simulating the full observable universe, the numerical data offers an ideal benchmark for the next generation of survey programs which will probe large cosmic volumes at high-redshift.

The realization of the first full universe simulation marks a remarkable achievement in the development of cosmological simulations. The improvements in memory and computing power are often described by the well-known ``Moore Law''. In cosmology this growth has resulted from
advancements in the efficiency of numerical algorithms as well as the increasing capabilities of supercomputing machines. These factors have been responsible for an exponential growth of the field, which has developed in parallel with major technological advancements.

\begin{figure}[h]
	\centering
	\includegraphics[scale=0.28]{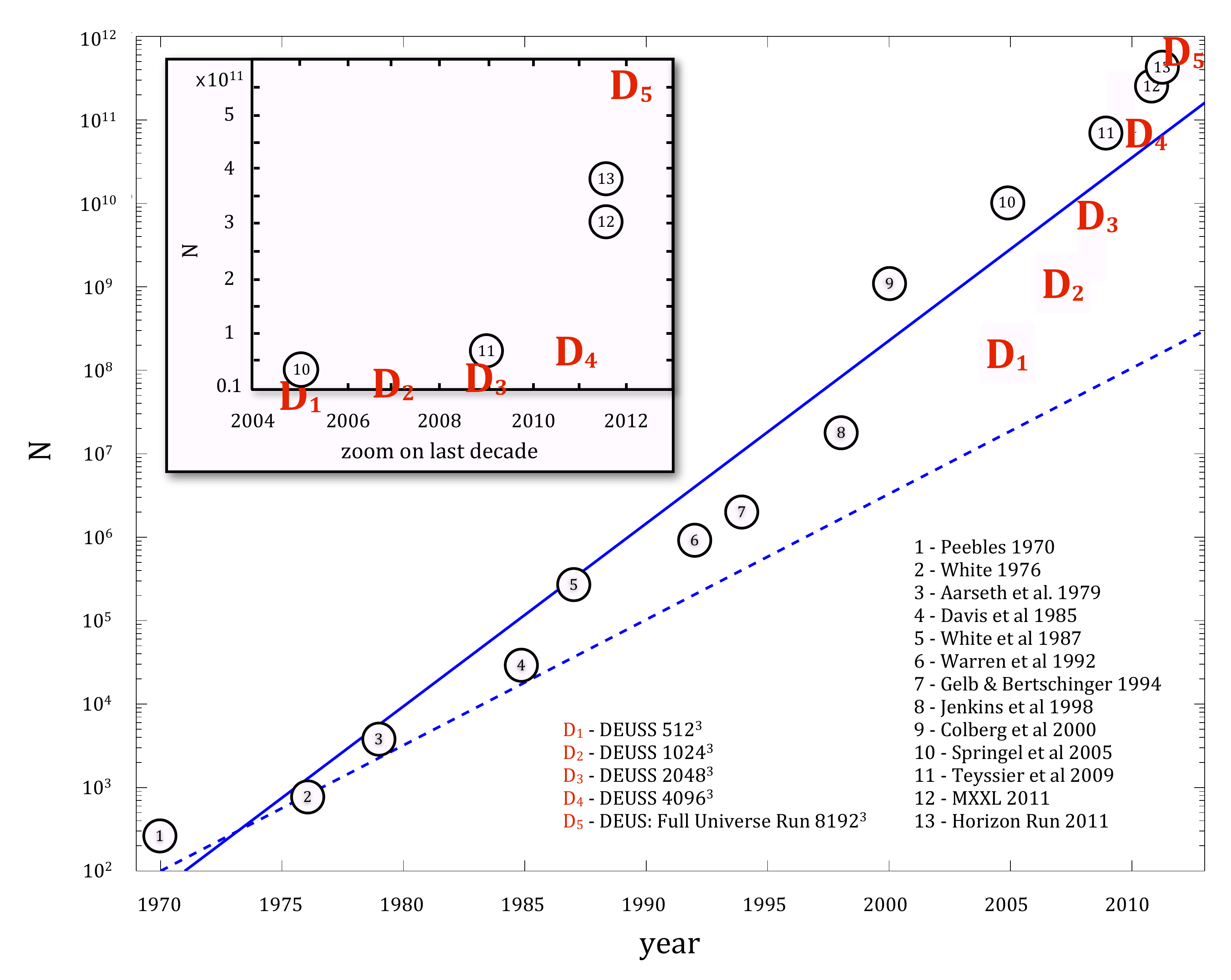}
	\caption{Evolution of the number of particles in N-body simulations versus time (years) from a non-exhaustive list of cosmological runs. 
D-symbols (red) are PM-AMR simulations made by our group. We can see the acceleration in performance occurred in the last decade especially for DEUS collaboration. 
Note in particular the position of the Millennium Run (10) \cite{Springel2005} with 10 billion particles and box size 
$500$ h$^{-1}$ Mpc; the Horizon 4$\pi$ (11) \cite{teyssier02} with 68 billion particles and box size $2000$ h$^{-1}$Mpc; the recent Millenium XXL Run (12) \cite{angulo12} with 303 billion particles and box size $3000$ h$^{-1}$Mpc and the Horizon Run (13) \cite{kim11} with 375 billion particles and box size $10800$ h$^{-1}$Mpc. The solid blue line is the mean evolution of the simulation size from Springel et al. (2005) and the dashed blue line is "Moore's Law" which shows a factor 2 increase every 18 months.}
	\label{moore}
\end{figure}

In Figure \ref{moore}, we plot a selection of cosmological simulations (the list is by no means exhaustive and is intended only to be illustrative of the evolution of the field). The first point on the graph is the historical simulation run by Jim Peebles at Princeton University in the early 70's with $300$ particles. The plot covers 40 years span including the latest high-resolution large-volume numerical simulations. First, we see that a "Moore"-like law with an increasing factor of $2$ every $18$ months underestimates the acceleration of state-of-the-art cosmological N-body simulations. The mean evolution of the simulation size is taken from Springel et al. 2005 \cite{Springel2005}. It is linear in logarithmic space, increasing by a factor of $10$ every $4.55$ years. In the smaller box we plot the  
trend from the DEUS Series \cite{DEUSS,rasera10} performed by our team during the last $7$ years with DEUS FUR at top of the ranked list.

\begin{figure}[h]
	\centering
	\includegraphics[scale=0.18]{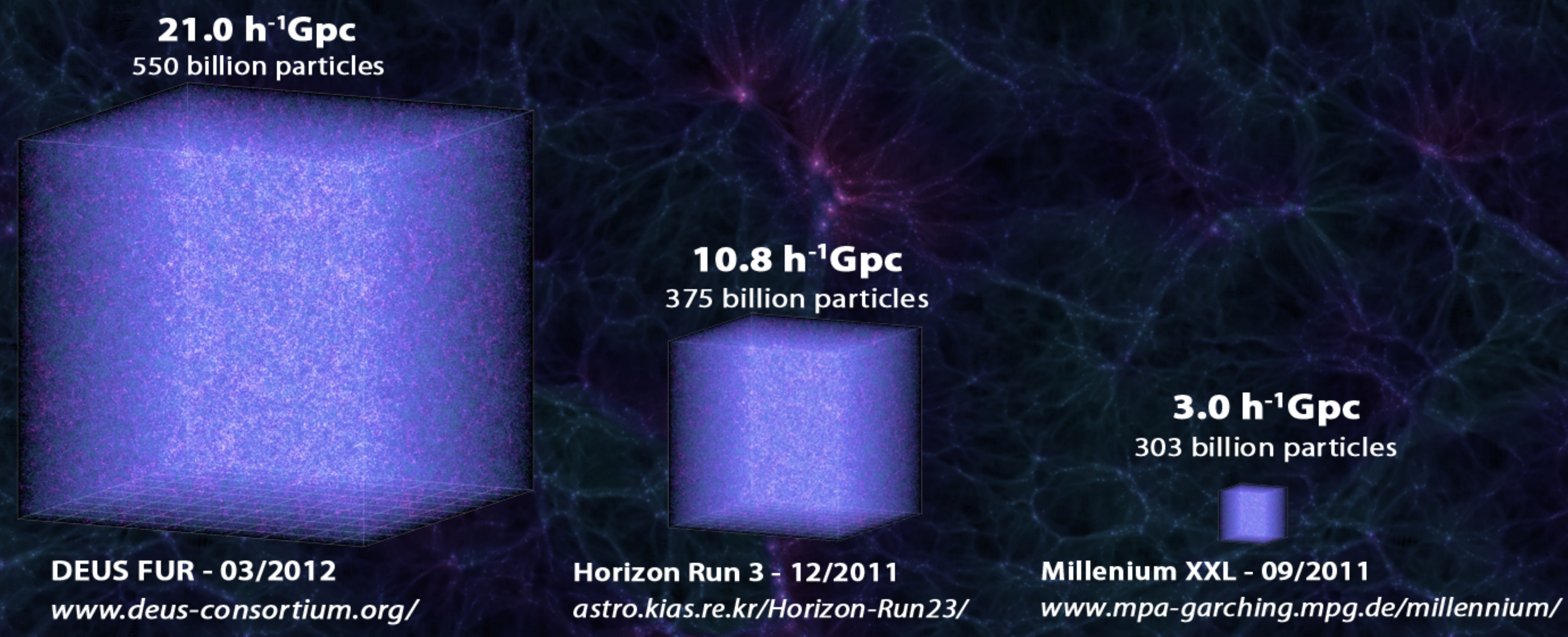}
	\caption{Comparison of the box size of DEUS Full Universe Run with the most performing numerical simulations in cosmology realized in the past few years. Pictures are in scale.}
	\label{box}
\end{figure}

We plot in Figure \ref{box} a visual comparison of box size of the three most recent cosmological ``grand-challenge'' simulations. The pictures are in scale.
We clearly see the impressive improvement of DEUS FUR with respect to the box size of Millennium XXL Run \cite{angulo12} and Horizon Run \cite{kim11}. 

DEUS FUR has the largest simulation box, nonetheless using an adaptive grid pattern we have been able to follow the gravitational dynamics over 6 orders of magnitude length-scale, from the size of the Milky Way to the size of the observable universe. Combining all numerical simulations performed by our group (DEUSS and DEUS FUR), we can trace the distribution of matter from a scale of less than one hundredth the size of our galaxy to the size of the full observable universe.

DEUS FUR has been realized through several steps which have required the use of the "AMA-DEUS"  (see Figure \ref{AMADEUS}). Firstly, the initial conditions are generated, then the 
gravitational evolution of the initial matter density field is computed, while the huge data volume generated during the run is properly organized, processed and stored. 
If we could back up all data characterizing a snapshot of DEUS FUR, i.e. the position and velocity of all particles, at all computing time this would amount to over 50 PB of data. 
This is obviously not possible (and would also be unusable), so we have developed a post-processing chain that can efficiently handle, considerably reduce and 
exploit the data generated during and immediately after the run. 

We describe in the following section the application AMA-DEUS.
 
\section{AMA-DEUS: Method and Application}

The dynamical evolution of Dark Matter particles under the effect of their mutual gravitational interaction in an expanding universe is described by Vlasov-Poisson equations which can be represented by a set of discrete point particles:
\begin{equation}
\label{eq:vlasov}
\frac{d\vec{r}_i}{dt} = \vec{v}_i \qquad \text{and} \qquad \frac{d\vec{v}_i}{dt} = -\vec{\triangledown}_r \phi
\end{equation}
with
\begin{equation}
\label{eq:poisson}
\Delta_r \phi = 4\pi G \rho \ .
\end{equation}
where $\vec{r}_i$ is the position, $\vec{v}_i$ is the velocity, $\phi$ is the potential and $\rho$ is the density.

This representation as an N-body system is a coarse grain approximation whose accuracy improves as the number of particles in the simulation box increases. Since we follow the growth of fluctuations in an expanding universe, the previous equations are re-written in term of comoving coordinates $\vec{x}$ with $\vec{r}= a(t) \vec{x}$, thus following expansion. The scale factor $a$, which appears explicitly, satisfies the Friedmann equation (homogeneous and isotropic solutions of Einstein equations). If we assume the cosmic matter content to consist of pressureless matter (composed of ordinary baryonic matter and cold dark matter), relativistic matter (photons and neutrinos) and a dark energy component, the expansion rate of the universe, defined in terms of the scale factor by the Hubble parameter $H =a'/a$ (a prime denotes a derivative with respect to the time), is given by the Friedmann equation 
\begin{equation}
\label{friedmann}
H^2=H_0^2\left[\frac{\Omega_{m}}{a^3}+\frac{\Omega_{r}}{a^4}+\Omega_{DE}(a)\right]
\end{equation}
with $H_0$ being the Hubble constant and $\Omega_i=\rho_i/\rho_c$ is the present value of the density parameters, where \mbox{$\rho_c=3H_0^2/(8\pi G)$} is the critical density and $\rho_i$ indicates the density contribution of pressureless matter, radiation and Dark Energy respectively. 

Several numerical methods have been developed to solve the gravitational dynamics of collisionless N-body systems. 
Grid-based N-body methods, such as the standard Particle-Mesh (PM)
whose robustness was demonstrated long ago, are particularly efficient \cite{Hockney1981}. In the PM method, the space where one computes the fields is discretized on a mesh, whereas the particles are evolved through the dynamics of these fields in continuous space. This method is decomposed into seven iterative steps:
\begin{enumerate}
\item compute the mass density $\rho$ on the mesh using an interpolation scheme from the position of particles. A Cloud-In-Cell (CIC) scheme is used;
\item compute the potential $\phi$ on the mesh using the field equation. In gravitational case, it is the Poisson equation which is solved by a multigrid method;
\item compute the acceleration on the mesh. A standard finite-difference approximation of the gradient is used;
\item compute each particle acceleration using the inverse interpolation scheme used in step $1$;
\item update each particle velocity according to its acceleration;
\item update each particle position according to its velocity;
\item check energy conservation equation and modification of the time-step.
\end{enumerate}

The computation of the gravitational dynamics of 550 billion particles on a range of scales going from the horizon (size of the observable universe) to the size of a galaxy implies several technical and numerical constraints in terms of computation time and memory management (optimization of memory usage and MPI communications are discussed below). 
There are also other equally challenging aspects, related to the generation of initial conditions, and most importantly to the management and organization of the huge volume of data generated during the run. The goal is to back-up only the most relevant data in a user friendly format available to users worldwide. These data consist of

\begin{itemize}
\item
$31$ "snapshots" corresponding to the backup positions, velocities and identifiers of all particles followed during the computation.
\item
$474$ "samples" for the backup of a $512$th of the simulation box at all computational coarse time-steps. We store not only the particles and their properties, but also AMR cells describing the gravitational field.
\item
$5$ light-cones constructed during the dynamical computation stored at all time-steps containing the particles and the AMR grid with the gravitational field in spherical shells around $5$ observers at different space-time points.
\end{itemize}

The complete chain of programs which includes the generation of initial conditions (MPGRAFIC-DEUS), the high-resolution computation of dynamical evolution of a very large number of particles (RAMSES-DEUS), the validation of computations (POWERGRID-DEUS), the post-processing workflow and data storage (PFOF-DEUS) are the software applications that have permitted the realization of DEUS FUR and are part of the "AMA-DEUS" which we describe next.

\subsection{Generating Initial Conditions}
\label{CI}

MPGRAFIC \cite{prunet2008} is a MPI code that generates an initial particle distribution (positions and velocities) from the primordial matter density fluctuations spectrum predicted in a given cosmological model. The power spectrum
inferred from the generated particle distribution is by construction exactly identical to that provided in input up to statistical fluctuations. This is realized by convoluting a white noise with the root of the input power spectrum. In addition to a few modifications implemented to achieve 
the desired level of resolution, we have modified the management structure of the input/output (I/O) necessary to handle a number of grid points at least equal to $8192^3$ . In the original version all processes write simultaneously in the same file. The splitting of the initial conditions into separate files and the implementation of a token system to prevent MPI processes from simultaneously accessing the file system have therefore been developed specifically for DEUS FUR and implemented in MPGRAFIC. Moreover, in the original version the FFT produces files in slices, while the dynamical code reads initial data in a domain decomposition along the Peano-Hilbert curve (see below). Hence, to overcome this limitation, the initial data are now directly sliced along the Peano-Hilbert curve in the code whatsoever the resolutions of the grid sizes and the number of particles used. To this end, we precompute the Peano-Hilbert curve using a large number of (very fast) processes instead of calculating it sequentially as in the original version. It is only then that the initial conditions are split along the Peano-Hilbert curve.

\subsection{N-Body Solver}
In the bottom-up hierarchical gravitational clustering picture, a large number of small clumps appear and merge progressively to form larger halos and filaments. Because of this, an adaptive resolution scheme is needed to improve the resolution of the N-body algorithm. 
This kind of dynamical resolution can be reached using an Adaptive Mesh Refinement (AMR) technique as in the RAMSES N-body solver originally developed by R. Teyssier \cite{teyssier02}. For our simulations, we have used an improved version of the original code that we briefly describe hereafter. 

The code is based on an AMR technique, with a tree-based data structure that allows recursive grid refinements on a cell-by-cell basis \cite{Kravtsov1997}. It takes advantage of both the speed of a mesh-based Poisson solver and the high-dynamical range and flexibility obtained with a tree structure. 
Particles are evolved using a particle-mesh solver \cite{Hockney1981, Alimi1993} , while the Poisson equation is solved with a multigrid method \cite{guillet11}. The refinement strategy follows a quasi-lagrangian approach where cells are divided by $8$ if their enclosed mass is multiplied by a given factor $\eta$. We do not impose any maximum level of refinement for the DEUS FUR simulations and let the code trigger as much refinement levels as needed. This allows us to reach a very high resolution from the scale of the Milky Way to that of the whole observable universe. AMR cells are recursively refined in RAMSES-DEUS if the number of particles in a cell exceed $\eta=14$. At the end of the DEUS FUR simulation $6$ levels of refinement have been reached for a total of two trillion AMR cells. As the coarse grid is $8192^3$, this corresponds to a formal resolution of $524288^3$ or 40 h$^{-1}$ kpc comoving spatial resolution in a 21~h$^{-1}$~Gpc computing box.

Each level of grid is evolved with its own time-step. A second-order midpoint scheme has been implemented, which reduces exactly to the usual second order leapfrog scheme \cite{Hockney1981} for constant time-steps. The time-step is determined independently for each level using standard stability constraints. The first constraint comes from the gravitational evolution which imposes that a variation $\Delta t^l$ of the time-step at level $l$ should be smaller than a fraction of the minimum free-fall time. The second constraint comes from particle dynamics within the AMR grid, this imposes that particles move only by a fraction of the local cell size. A third constraint is imposed on the time-step by specifying that the expansion factor $a$ should not vary more than $10 \%$ over one time-step. This constraint is active only at early times during the linear regime of gravitational clustering. Finally, the actual modification of the time-step is equal to the minimum imposed by the three previous constraints.

RAMSES has been parallelized using a dynamical domain decomposition based on the Peano-Hilbert space-filling curve. This is an important feature of the code for high-resolution runs with a high level of clustering. 

In order to implement the cosmological evolution of the different models we have implemented all cosmological routines such as to make them model-independent. These are now inputs with pre-computed numerical tables of cosmological quantities needed to fully specify the homogeneous cosmological model evolution. 

\subsection{Post Processing Workflow}
Our post-processing workflow has three main goals: validate the simulation run on-flight by measuring and testing physical quantities, significantly reduce the data generated during the run, provide a preliminary analysis of the data and organize them in a user-friendly manner.

For every simulation snapshot we compute the inverse Fourier transform of the density field and the power spectrum using the code POWERGRID-DEUS.
Because of the large number of particles in the simulation box (550 billions), to attein a sufficiently high spatial resolution (FFT on a $16384^3$ grid) 
this analysis can be performed only using a parallel version of the code running  on 16384 cores. 
The run is then verified by comparing the inferred power spectrum against the linear theory prediction. We also verify 
that all conservation laws such as the energy conservation are well satisfied during the run.

Once the data have passed the acceptance test, they are further processed by the application PFOF-DEUS. This is an MPI code that has been
entirely developped within the DEUS consortium for DEUS and DEUS FUR. It consists of 4 essential tools: PFOF-Slicer (data slicer in cubes), PFOF-Finder (FOF halo finder), PFOF-prop (properties of cubes and halos), PFOF-multi (compression and data archiving). First, PFOF-Slicer runs to perform a cubic splitting of the RAMSES-DEUS Hilbert space-filling curve, making the data much more user-friendly, then PFOF-Finder detects dark matter structures in the underlying matter field using a percolation method (Friends-of-Friends algorithm). In addition PFOF-DEUS builds standardized outputs: 
sub-cubes (.cube) (limited region of computing box), halos (singular object extracted from the matter field) with their particles (.halo) and their center of mass (.cdm), thus providing a survey of massive halos of the RAMSES-DEUS outputs (snapshots and coarse outputs). RAMSES-DEUS output files are ordered along the Hilbert curve, hence the cubic splitting is achieved through a double reading/answering the question: where are the particles belonging to one sub-cube in the Hilbert curve? This algorithm has the advantage of avoiding large communications. The outputs are single-precision and do not contain information about gravity AMR cells. This procedure results in a gain of a factor 7 in disk usage.

Halos and their particle constituents, as well as the cubes are stored only every two snapshots (odd-numbered snapshots). For the remaining even-numbered snapshots we define halos more loosely using a larger value of the percolation parameter. This allows us to store particles in larger regions where the halos have formed, so that a more precise detection of the halos which are in these odd snapshots can be performed at level of post-treatment analysis. This first data reduction process allows us to reduce the amount of used disk space by a factor of about 300 compared with the complete storage of all data for all snapshots. This procedure is applied to the RAMSES-DEUS data all over the simulation box as well as the sample data which are saved at each computing time-step of the simulation. The number of MPI tasks necessary to run this procedure amounts to 64 for a "sample" and 32768 for a complete snapshot of the entire simulation box.

Using the halos detected in snapshots, PFOF-prop compute physical quantities on both cube and halo files: cube\_properties uses only one core, it reads the header of all cube files and computes a rough estimation of various quantities such as cube density; halos\_properties uses the same number of cores as the FoF-Slicer (i.e. between 64 and 32768 cores) and computes fine observables on every dark matter structures such as the virial radius, the kinetic energy, the potential energy, the inertia tensor, etc. 

The last step of the workflow consisting in backing up the data on an independent file system. As the number of files handled by this file system is low, it is necessary to use of a compression technique. PFOF-multi performs a tar-like operation on several cores by reading the original PFOF-Slicer output files, compressing them into a smaller file (usually 64 files are compressed into one) and writing it on the storage file system. The number of MPI tasks used by this routine varies from 1 to~512.

\section{CURIE Petascale supercomputer}
\begin{figure}[h]
	\centering
	\includegraphics[scale=0.5]{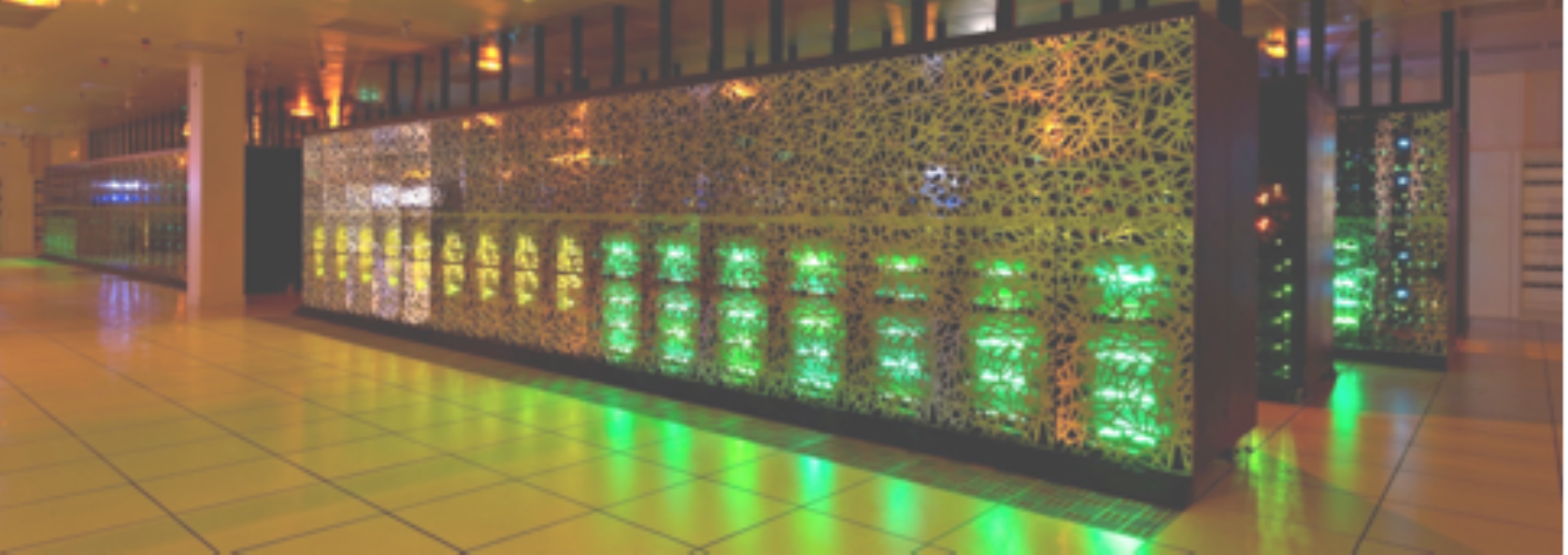}
	\caption{Picture of GENCI CURIE supercomputer.}
	\label{curie}
\end{figure}

\begin{figure}[h]
	\centering
	\includegraphics[scale=0.27]{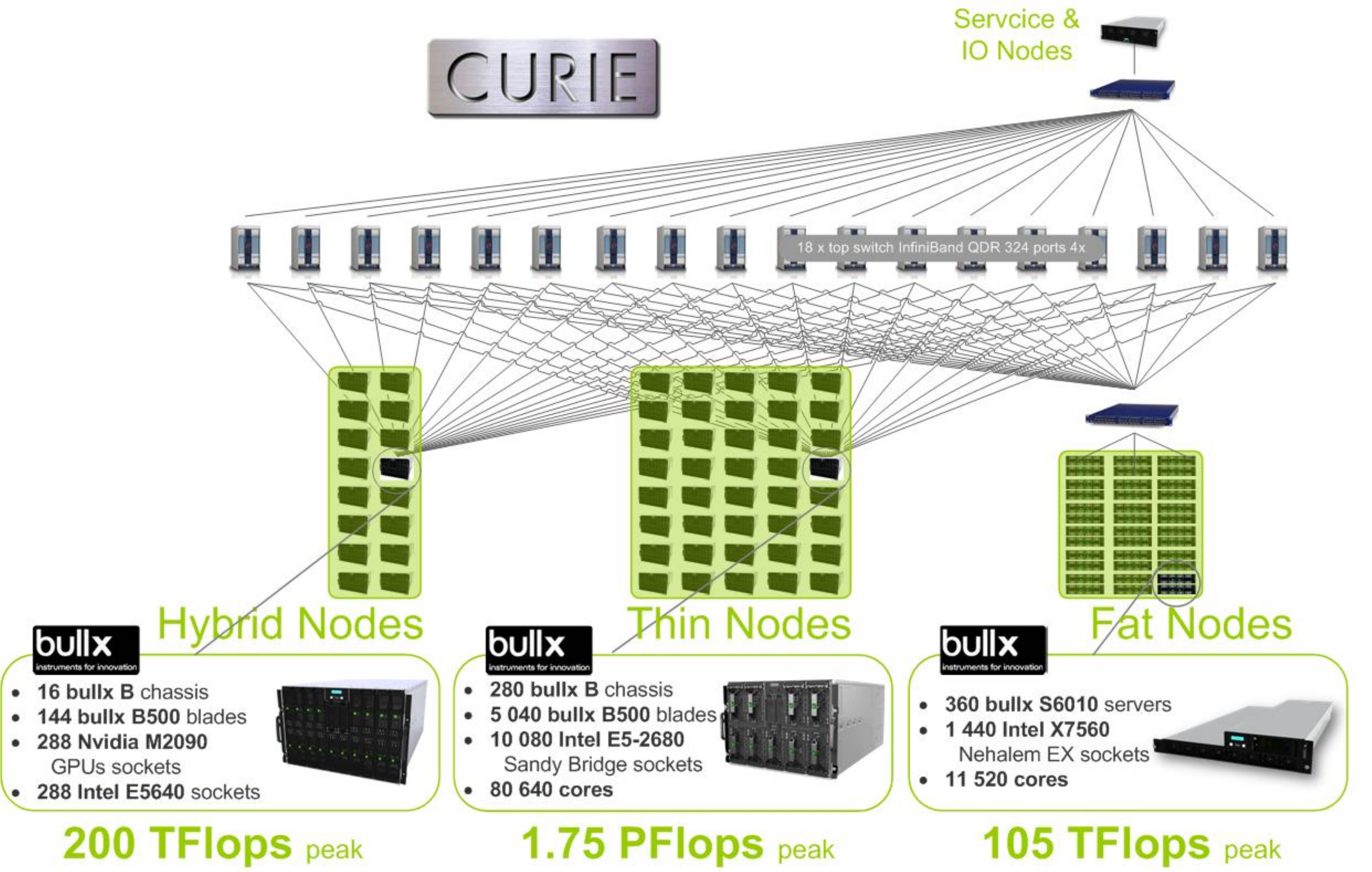}
	\caption{CURIE full architecture.}
	\label{curiecluster}
\end{figure}

CURIE is the second PRACE Tier0 petascale system provisionned by GENCI and operated at TGCC (Tr\`es Grand Centre de Calcul du CEA) near Paris. 
It is one of the ten machines in the TOP500\footnote{www.top500.org} with a 2 Petaflops peak BULL Bullx cluster with a modular architecure based on three complementary x86 partitions: 
CURIE is a Bull Bullx cluster \ref{curiecluster} encompassing three different types of compute nodes:
\begin{itemize}
\item	
$90$ large nodes with $16$ Intel Nehalem-EX processors at $2.26$ GHz ($8$ cores each), $512$ GB of DDR3 memory ($4$ GB/core) and $4$ ConnectX QDR Infiniband attachment : each node actually consists of four standard $4$ processors servers interconnected via a so-called Bull Coherency Switch (BCS) to make up super-nodes
\item 
$5040$ Bullx B510 thin nodes with $2$ Intel SandyBridge processors at $2.7$ GHz ($8$ cores each), $64$ GB of DDR3 memory ($4$ GB/core), SSD local disk and one ConnectX QDR Infiniband attachment.
\item 
$144$ Bullx B505 hybrid nodes with $2$ Intel Westmere (4 core) 2.67 Ghz and $2$ Nvidia M2090 GPUs each
\end{itemize}
All these compute nodes are connected by a full fat tree QDR Infiniband network.

The different types of nodes may match different application requirements. Some preliminary developments of our project have been performed on the large memory nodes but for running our DEUS FUR simulations we used the thin nodes partition in order to locate the dataset into more than 300 TB of memory. 

CURIE has a modular and balanced architecture, not only characterized by its own raw compute power but also by a huge memory footprint (more than 360 TB of distributed memory) and I/O capabilities. CURIE has a 2 level LUSTRE file system with a first private 5 PB internal level accessed at 150 GB/s and a second shared level of around 8 PB at 100 GB/s.

\begin{figure}[h]
	\centering
	\includegraphics[scale=0.26]{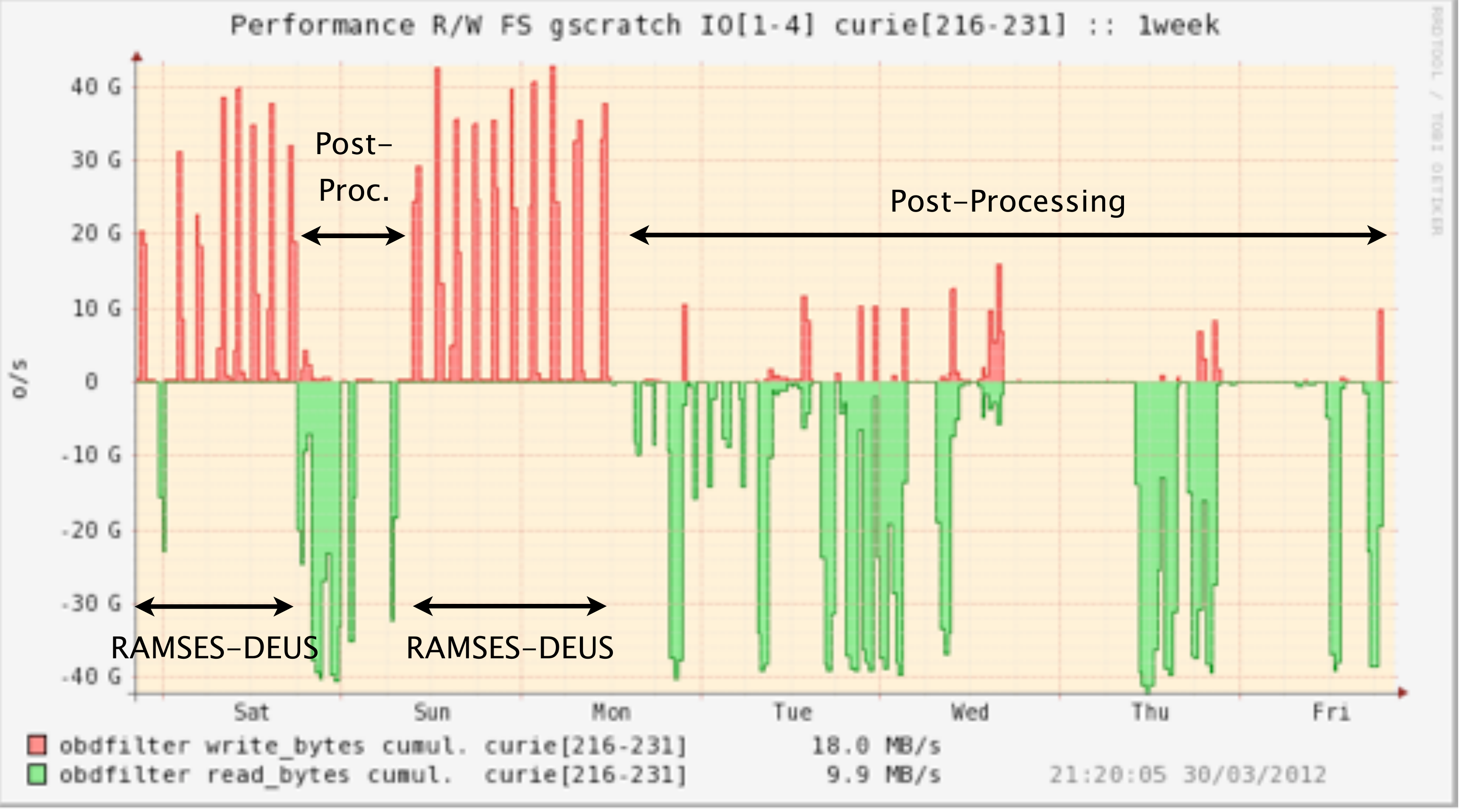}
	\caption{I/O statistics during DEUS great challenge.}
	\label{debit}
\end{figure}

A part of the first level private LUSTRE parallel file system has been dedicated to DEUS grand challenge: 1.7 PB with a 60~GB/s bandwidth, which was used at almost full speed (figure \ref{debit}: more than 40 GB/s writing during half an hour (red) and the same reading speed (green)).

\section{AMA-DEUS: Numerical Challenge and Numerical Performances}

The realization of DEUS FUR has been possible thanks to the optimization of all codes used for the project. These optimizations have been mandatory, not only during the preparatory phase of the calculations, but also during the computational run to obtain the best performance of the supercomputer CURIE. In the following paragraphs we describe in detail these optimizations, the various adjustments that their implementation has required and results obtained on various aspects of our application: computation time, communication, memory usage and I/O.

The system configuration that have been chosen to follow the dynamical evolution 
of $8192^3$ particles with RAMSES-DEUS includes $38 016$ MPI tasks distributed on $4 752$ nodes of CURIE-Thin, each with $8$ GB of memory for a total of approximately $300$ TB of RAM. In practice, during the computation, $3$ Po of data have been written and each MPI task has used an average of $6$ GB of memory. Before the ``production'' phase, three other simulations in weak-scaling configuration have been performed on CURIE-Thin to test the scalability of our codes and adjust the optimization parameters accordingly. In the following paragraphs we compare in details the characteristics of these test runs which are listed in Table I.

\begin{table}[!h]
\label{tabletest}
\begin{center}\caption{Characteristics of the test run simulations in weak-scaling configuration.}
\begin{tabular}{|c|c|c|c|}
 \hline
 Particles & MPI Task Number & MPI Task Memory & Nodes Number \\
 \hline
  $1024^3$ & $74$ & $8$Go & $10$ \\
  $2048^3$ & $594$ & $8$Go & $75$ \\
  $4096^3$ & $4752$ & $8$Go & $594$ \\
  $8192^3$ & $38016$ & $8$Go & $4752$ \\
  \hline
\end{tabular}
\end{center}
\label{Table}
\end{table}

\subsection{Initial Conditions}

Splitting the initial conditions (see section \ref{CI}) along the Peano-Hilbert curve is one of the many aspects necessary to the realization of DEUS FUR. The generation of initial conditions in the sliced configuration for $8192^3$ particles takes 41 minutes on 2048 MPI Tasks with 4 GB memory per task. This time includes the time to write $\sim 8$ TB of data as well as reloading them using a conservative system of tokens (one process among 64 is acting as I/O node) for reading and writing 250 MB files. For initial conditions with $4096^3$ particles, the run time is 38 minutes on 256 MPI tasks. 

Concerning the calculation of the Hilbert-Peano curve our application uses 38016 MPI tasks instead of the single one implemented in the original version of the code. This reduces by a huge factor the running time of the application (several thousands, the exact number has not been preserved), thus making this operation feasible. Uploading $\sim 8$ TB of initial condition data on the code RAMSES-DEUS, including the time for the initialization phase, takes about 23 minutes to be confronted with 1-2 days if we did not split the initial conditions.

\subsection{Computing-time and weak-scaling study of RAMSES-DEUS}

\begin{figure}
\centering
 \includegraphics[scale=0.32]{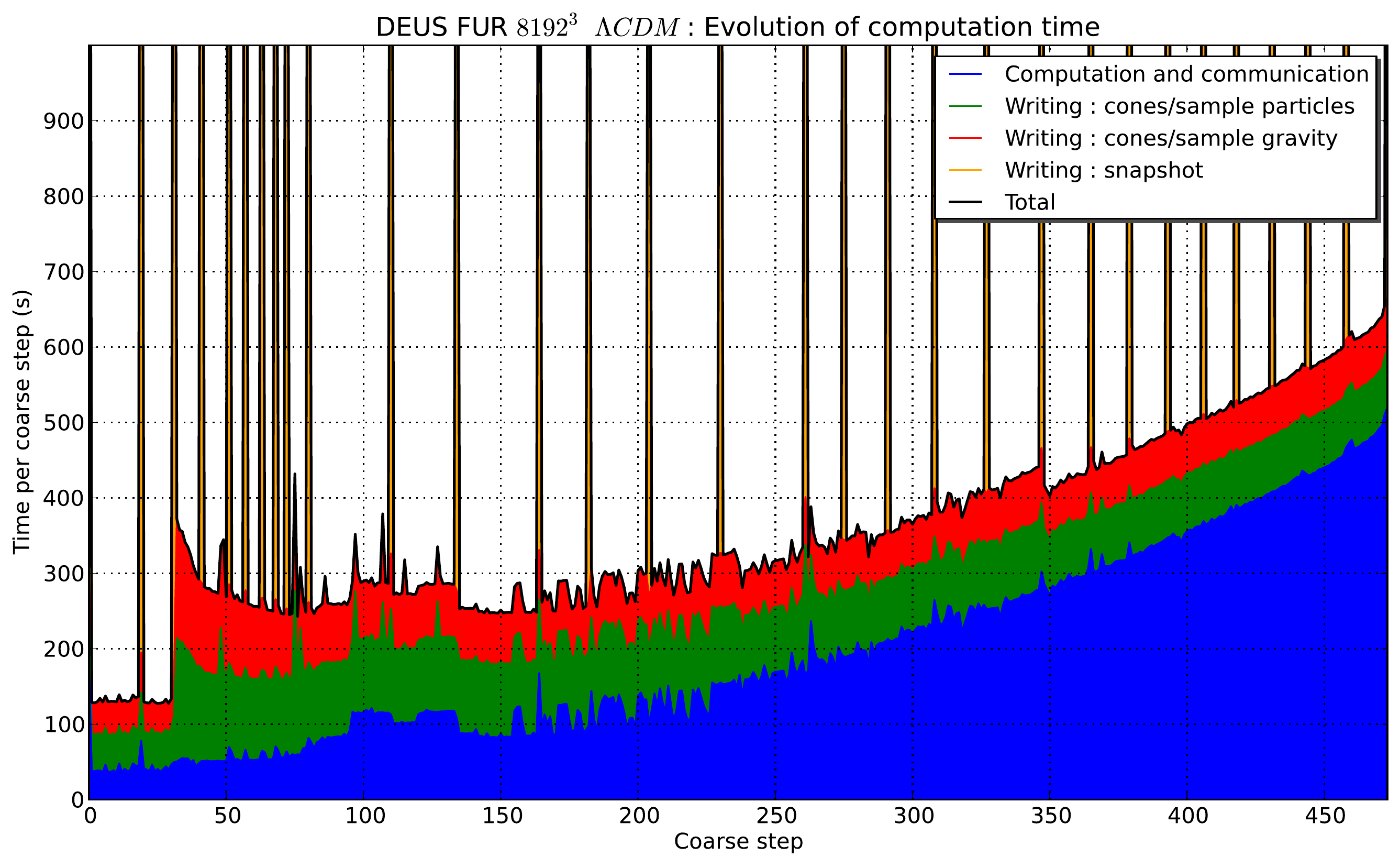}\\ 
\caption{Various contributions to the total elapsed time as a function of coarse time-step number. CPU time is shown as blue, time for writing particles and gravity coarse output in green and red, time for writing snapshots in orange (each peak is between 1500 and 3000s). The discontinuity near time-step 30 corresponds to the beginning of the writing of the light-cones. Before this time-step, only the "sample" is written. CPU time increases because the number of cells increase, time for coarse output decreases because the amount of particles inside the light-cones becomes smaller.}
\label{timings}
\end{figure}

\begin{figure}[b]
\centering
 \includegraphics[scale=0.33]{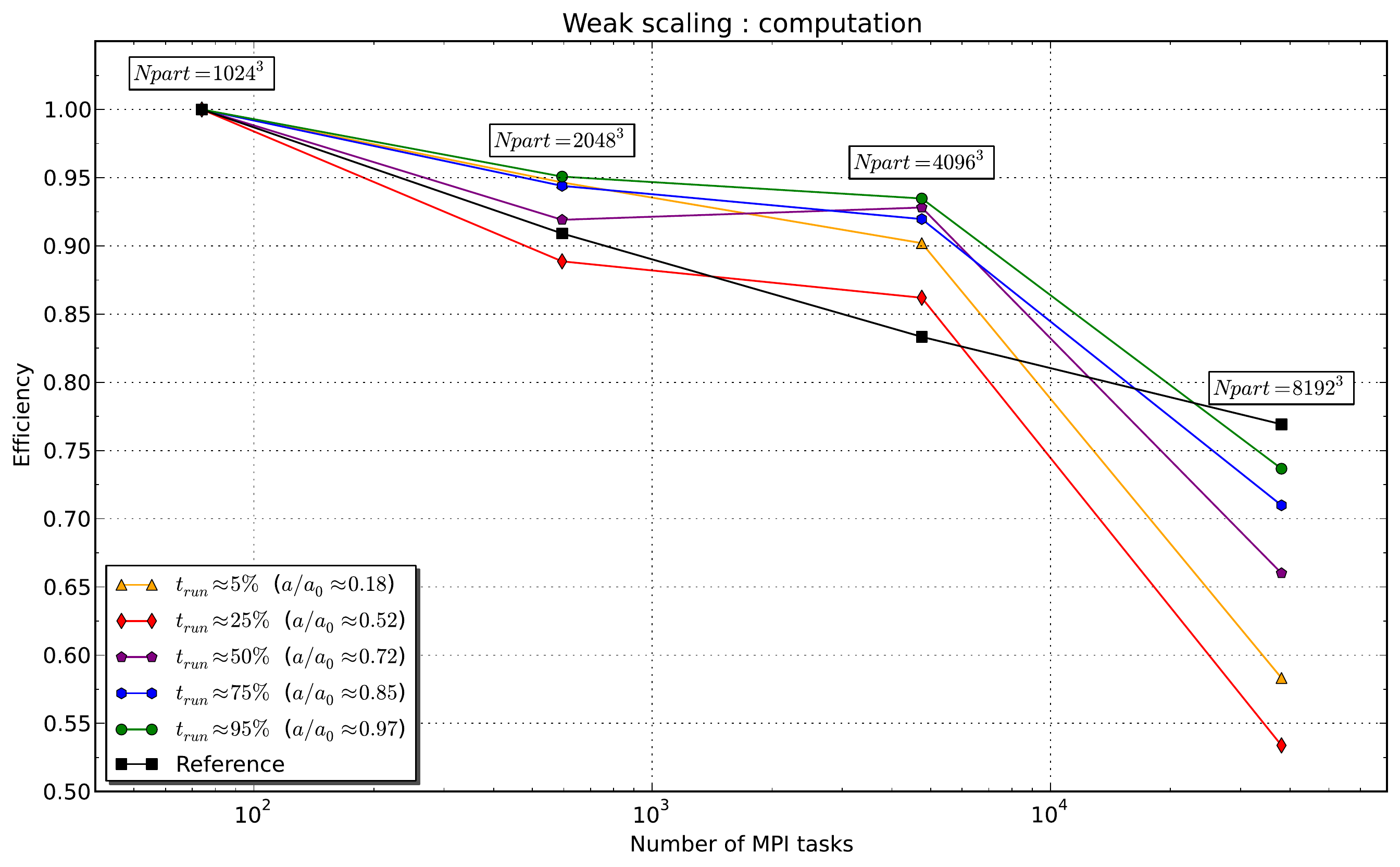}\\
\caption{Efficiency of N-body/Poisson solver as a function of the number of MPI tasks in a weak-scaling configuration. The reference corresponds to 74 MPI tasks. The efficiency is shown at the beginning of the run (yellow), at 1/4th (red), half (purple), 3/4th (blue) and at the end of the run (green). . The efficiency is first of the order of $60\%$, it falls to about $55\%$ during a short time when the first refinements are triggered and finally it increases to $ 75\%$. Multigrid acceleration allows us to reach higher efficiencies comparatively to the efficiency of an ideal PM-FFT code in black.}
\label{scaling}
\end{figure}

Other crucial aspects for the successful realization of DEUS FUR concern the optimization of the effective computational time of the particle dynamics and the I/O performance through the LUSTRE file system (see Figure~\ref{timings}). As structures in the simulation are formed, the AMR tree creates new refinements, leading to a gradual increase of the time required to calculate each time-step. This evolves from $50$ seconds to nearly $500$ seconds during the calculation. On the other hand, the writing time of the data particles and the gravitational field of the light-cones as well as the sample, all of which occurs at each computational time-step, decreases as the amount of light-cone data gradually decreases during the run. 
This time, however, remains comparable in magnitude to the computational one. Hence, to improve the former a dynamic system of tokens has been implemented in the code (see \ref{sec:ramsesIO}). Finally, the run is divided in $31$ phases of intense writing in which about $60$ TB to $120$ TB are written on the disk. 
The stability of the file system is critical during these phases that can take from $1500$ seconds at the beginning of the run to $3200$ seconds at the end.

Let us now focus on the performance of RAMSES-DEUS in the ``weak-scaling'' configuration in terms of computation time and MPI communications. The test runs where performed using $1024^3$, $2048^3$ and $4096^3$ particles (see table I). These were used in combination with the $8192^3$ particle simulation to optimize the relevant parts of our codes and improve their efficiency. This has ultimately led to the performance illustrated in Figure ~\ref{scaling}. 
We can see that up to 4752 MPI tasks ($4096^3$ particles simulation), RAMSES-DEUS reaches an efficiency of about $90\%$ compared to the reference simulation with 74 MPI processes ($1024^3$ particles). For comparison, this exceeds the theoretical efficiency of a standard ``Particle-Mesh'' solver (without refinement) which solves the Poisson equation through FFT. This is due to the solver of the original version of the code RAMSES which is based on a multi-grid method \cite{guillet11}. In the case with $38016$ MPI tasks, which is equivalent to that of the simulation DEUS FUR, the efficiency slightly drops. However, this decrement depends on the computational phase of the system dynamics. At the beginning, when there is not refinement, the efficienty is about $60\%$, then it falls to $55\%$ when the first level of refinement are triggered. Nevertheless, the decline remains limited to about $15\%$ of the total computation time. After this phase, the efficiency increases all along the simulation run when the total number of cells becomes very large. It remains above $65\%$ for more than half of the run, reaching up to $75\%$ efficiency, which is nearly the ideal efficiency of a PM-FFT solver.
 
To improve scalability, different communication schemes have been implemented in addition to the default method already present in the original version of RAMSES \cite{teyssier02} and based on asynchronous MPI operations (Isend/Irecv). Synchronous communications (Send/Recv, Bcast) have been implemented for certain levels of refinements (particularly the coarser ones). Such optimisations have proven to be highly dependent on the MPI library as well as the IB topology.
This has required the implementation of specific tuning of the system by BULL HPC experts during the preparatory phase.

\subsection{Memory usage}

\begin{figure}
\centering
 \includegraphics[scale=0.32]{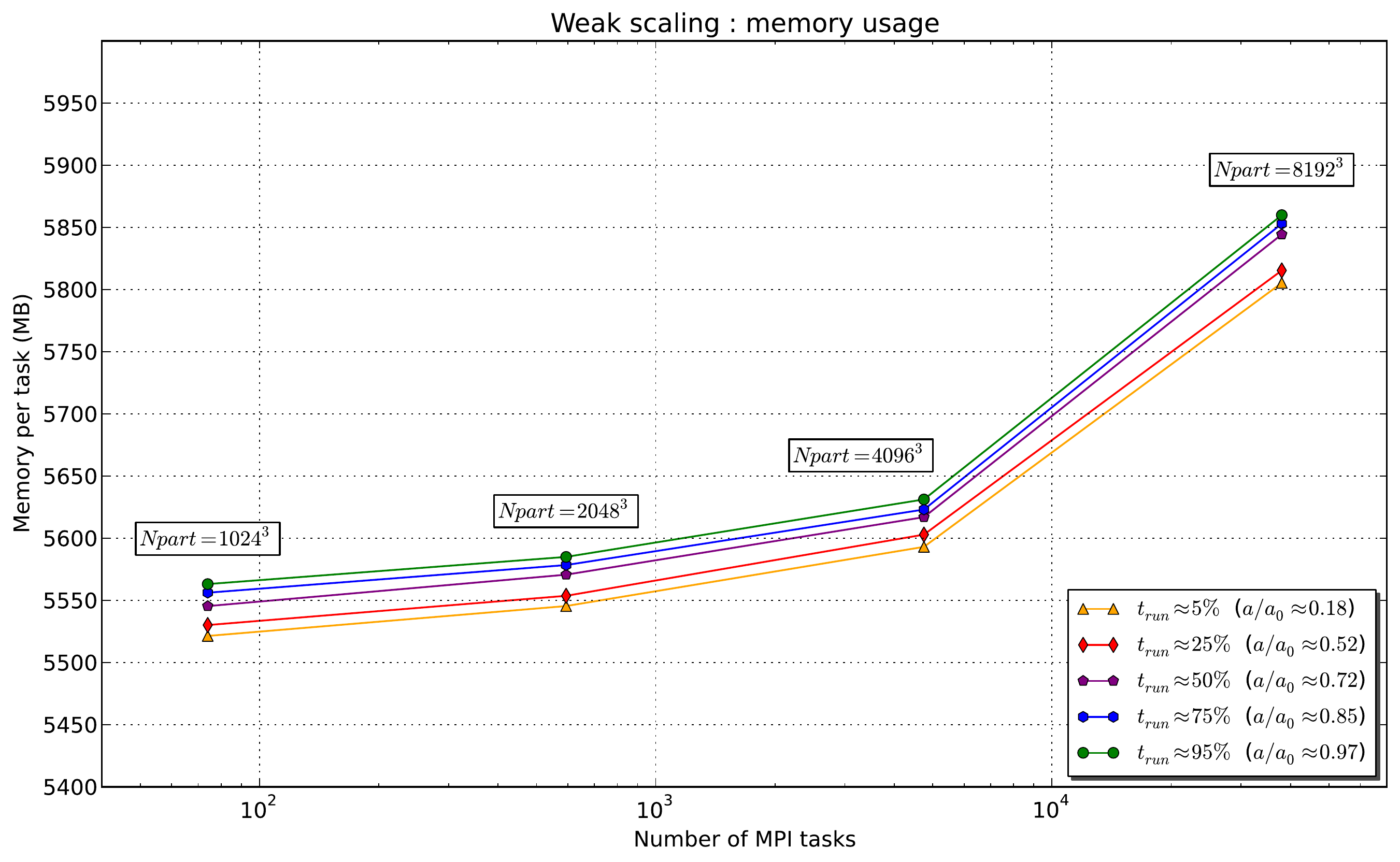}\\ 
\caption{Mean memory usage as a function of number of MPI tasks in a weak-scaling configuration: beginning of the run (orange), at 1/4th (red), half (purple), 3/4th (blue) and at the end of the run (green). The control of memory usage (including the one from MPI buffers) achieved at the 5\% level was a key point for the success of the run.}
\label{memory}
\end{figure}

Memory management has been one of the major challenges of the simulation. This is because in order to realize the simulation the computation has to satisfy the absolute condition of using less than $64$ GB per node, i.e. less than $8$ GB per MPI process. Any ``swap'' generated by a local excess of memory leads to a prohibitive execution time.

Thus, to evaluate the memory used in the weak-scaling configurations, a set of preparatory tests were designed before the availability of CURIE ``Thin'' on the fat nodes partition of CURIE. The goal of these tests has been to determine the best parametrization of the triggering levels of the adaptive mesh refinement from a physical point of view (essentially the final choice of the threshold sets the refinement to $14$ particles per cell) for different scenarios of memory usage. The execution of these tests has represented a milestone toward the realization of the project. These have allowed us to
to maximize the accuracy of the calculation of the dynamical evolution of the particles given the available amount of memory on each node of CURIE Thin.

The size of arrays, which is proportional to the number of MPI tasks, can become problematic when using tens of thousands of cores. We have found such memory problem in previous DEUS runs that used $25000$ MPI processes on a Blue Gene/P supercomputer \cite{IDRIS} (IDRIS) in which every core has only $512$ MB of local RAM. The problem has been identified in the arrays of pointers whose size is proportional to the number of MPI tasks multiplied by the number of refinement levels. Hence, an important work has been done to use intermediate arrays of pointers allocated dynamically if necessary, thus leading to a reduction of about a factor of 3 in the amount of memory used by RAMSES-DEUS.

The communication ``buffers'' created by the MPI libraries have been critical in terms of memory usage. This has emerged during the testing phase of 
CURIE-Thin. To identify the problem we have used the application ``Valgrind'' \cite{VALGRIND} to monitor the memory used by RAMSES-DEUS.
During the initialization phase, $9$ processes communicate with all $38016$ processes using the original communication scheme of the RAMSES code which is based on point-to-point communications. Because of these, 9 tasks use up to $16$ GB of memory when the MPI libraries are configured for the optimal execution speed (via OpenMPI options), thus altering the normal run of the simulation till its complete crash. The use of OpenMPI parameters with low memory footprint has solved the problem, however this resulted in lockups of the code. An ``intermediate'' parametrization has been finally chosen, this includes a relaxation of the option scheme of the generic point-to-point communication in favor of BCAST instructions which have reduced the maximum memory used per processes by about $35\%$, thus limiting the memory usage of the 9 processes considered to $8$ GB.

All these optimizations have led to an average use of memory per process illustrated in Figure \ref{memory}. The majority of arrays are allocated during the initialization phase of the code, the average memory used increases very little during the run and remains below $6$ GB per process. We can see that there is a good memory scalability, since increasing the number of MPI tasks from 74 to 38016 increase the memory usage by $5\%$ only.

\subsection{Parallel I/O}
\label{sec:ramsesIO}
In order to fully appreciate the complexity of the large data volume to be handled, let us 
remind the various types of RAMSES-DEUS outputs. The largest outputs ($\sim$100 TB) 
are the so called ``snapshots'' produced $31$ times, while smaller ``coarse outputs'' are produced at every coarse time-step during the computation. The latter are divided into two categories: a ``sample'' occupying a volume $1/512$th of the full computational box and a typical size of 140 GB; 
shells of a ``light-cone'' with a size varying from 600 GB to 5 MB. Using MPI blocking instructions such as MPI\_Send/MPI\_Recv, the number of concomitant tasks writing the data has been fixed such as to saturate the bandwidth of the system. 594 simultaneous writings are allowed in the case of snapshots, whereas in the case of the samples all tasks can write at the same time. The large variation in the size of shell outputs requires the use of an adaptive token system. This is set such that at each time-step the ratio of the volume of the overall box to the shell volume defines the number of concomitant writings. 

\begin{figure}[b]
\centering
\includegraphics[scale=0.32]{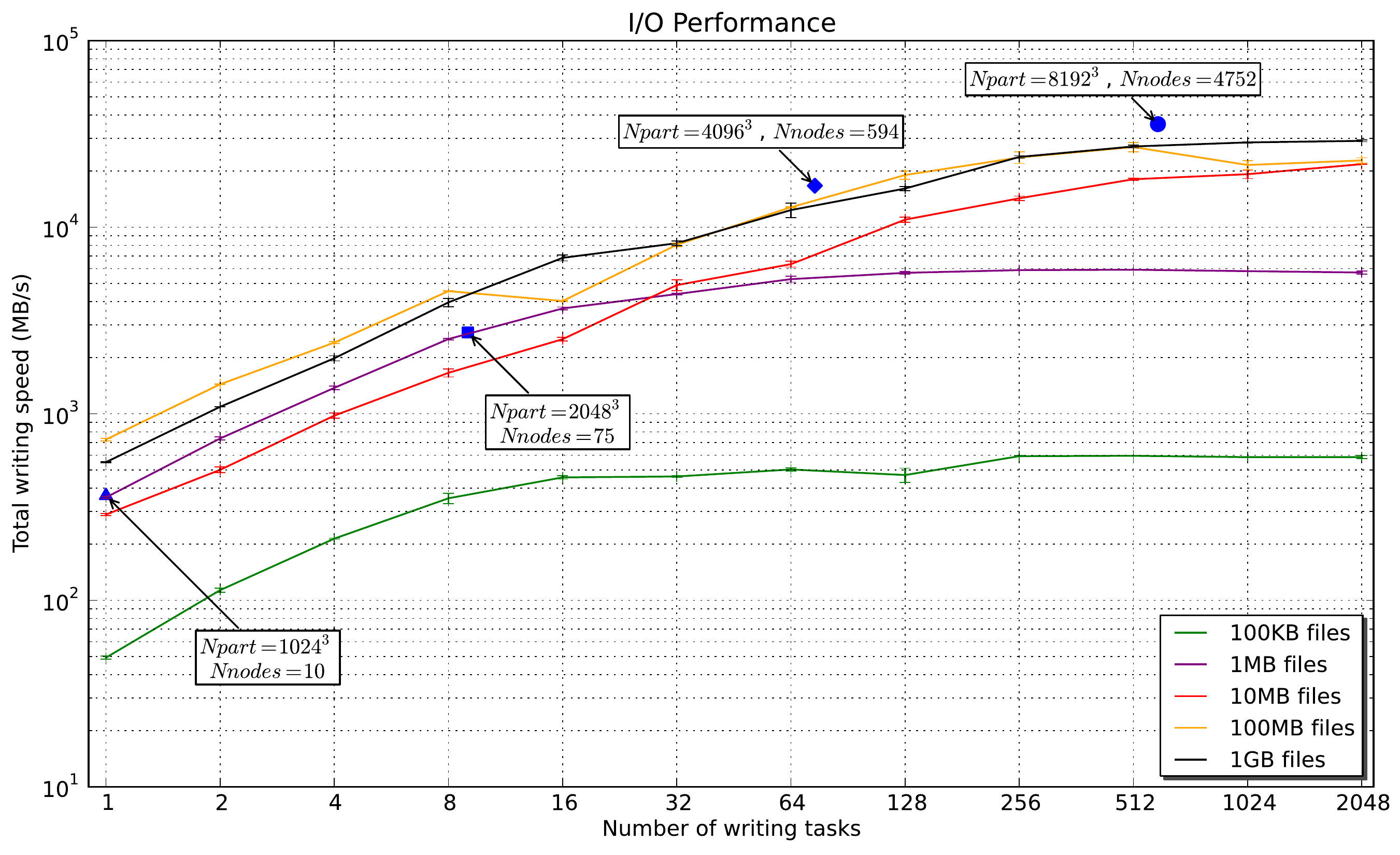}\\ 
\caption{Writing speed as a function of number of MPI tasks: green is for 100~KB files, purple 1~MB, red 10~MB, yellow 100~MB, black 1~GB like in our run. This is measured using a benchmark on less than 128 nodes. Blue points correspond to the average writing speed during the whole Full Universe Run (4752 nodes/38016 MPI tasks), and the weak scaling simulations. The token system was tuned to saturate the bandwidth allocated for our simulations.}
\label{io}
\end{figure}

Before running the simulation, we have run I/O benchmarks whose results are shown in Figure \ref{io}. As one can see the optimal size for files is above 100~MB, the size of data files generated during the dynamical computation is of order 1~GB. Moreover, the writing speed saturates above 512 tasks. Therefore, we have chosen 594 simultaneous writings (corresponding to 1 task every 64). We note that the average performance during the run was almost 40~GB/s which is above the result of the benchmark. This is because the benchmark was run only on a small fraction of the supercomputer cores. It appears that more than 128 nodes must be used in order to aggregate the full I/O bandwidth of the LUSTRE file system.

A final and very important remark concerns the concomitant access of all tasks to the same file during the reading (for instance input parameters). This process may lead to instabilities in the LUSTRE file system. We have therefore used broadcast MPI instruction: a single MPI task reads the parameters files, pack them in dedicated MPI structures and broadcasts them to other tasks.

\subsection{Post Processing Workflow}

Similarly to the modifications of the I/O management structure in RAMSES-DEUS, 
we have implemented a token system in POWERGRID-DEUS (8192 cores) and PFOF-Slicer (up to 32768 MPI tasks).

For POWERGRID-DEUS, since the Fourier transform is faster along planes than with a cubic splitting, we have implemented a data reading from RAMSES-DEUS and a redistribution of these data along planes via MPI communications. The parallel computation of the density has been performed using a CIC ("Cloud In Cells") scheme. The power spectrum is deduced from the inverse Fourier transform of the density. All the RAMSES-DEUS data need to be processed by POWERGRID-DEUS on a $16384^3$ grid (4000 billion computing points). The running time is limited to 1.2 hours with $8192$ MPI tasks and $4.2$ GB of memory per task. 

The optimization of PFOF-Slicer has been designed to minimize the reading of the huge amount of data from RAMSES-DEUS to less than 2 hours with less than 100 bytes per particles. The initial cubic decomposition algorithm of particles distributed along the Peano-Hilbert curve has used a scheme with a low number of readings followed by numerous MPI communications. However, the time needed for the communications rapidly increases with the number of processors causing the processing time to explode. Another problem with this method is the doubling of the memory required for the splitting. This is because in order to bring together the particles in a cube associated with a given MPI process, a temporary table is created with a size equal to twice the number of particles. To avoid these two major issues, we have developed a method which precludes all communications between processors but instead uses a large number of readings. This solution consists of two phases. In a first phase the application reads only the number of particles per MPI process and their local distribution map. 
Using RMA (Remote Memory Access) MPI2 instructions, the complete map is recovered instantly. In a second reading phase, using the distribution map, the RAMSES-DEUS data are written directly into final variables by reading only the necessary files. This new implementation allows the use of PFOF-SLICER on 32768 cores and a gain in speed by a factor of 2 on the cubic splitting. In this way, a snapshot can be processed in only 2 hours.

\begin{figure}[h]
\centering
\includegraphics[scale=0.32]{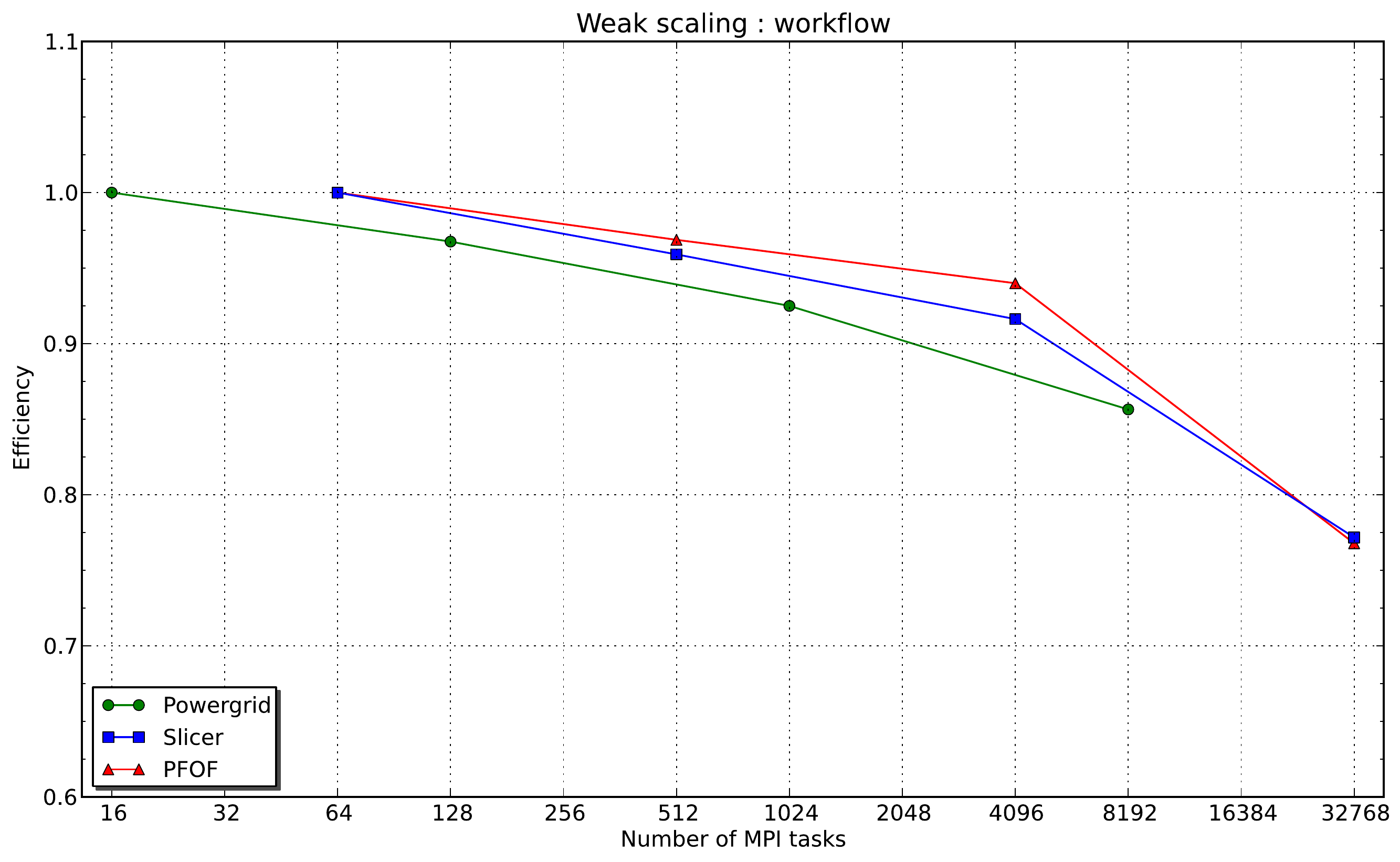}\\ 
\caption{Efficiency of programs developed to process data generated during the DEUS FUR simulation as a function of the number of MPI tasks. The efficiency obtained is satisfactory: The computation of the power spectrum for validating the results of RAMSES-DEUS using the application POWERGRID-DEUS corresponds to green curve. The redistribution of these data along the Peano-Hilbert curve in cubic splitting by the PFOF-Slicer application is the blue curve. The detection of massive halos by the percolation algorithm "Friends of Friends" \cite{knebe11} is obtained using PFOF-DEUS, its efficiency is represented by the red curve.}
\label{io2}
\end{figure}

Backup of the data on the disk by ``tar'' and ``copy'' instructions has turned out to be an option that takes too long due to the amount of data that is necessary to transfer (several months for all data). To solve the problem we have developped the application PFOF-Multi for the PFOF-Slicer code. This option enables the concatenation of small files (500 MB) in files with sizes between 10 and 100 GB and the writing directly to the backup system. This strategy has turned out to be very effective with a transfer rate greater than 5 GB/s even in the presence of many other users. 

The efficiency in ``weak scaling'' configuration of some aspects of data processing is shown in Figure \ref {io2}. I/O performance are presented in Figure \ref{debit}.


\section{Preliminary Science Results and Perspectve}
A preliminary analysis of the DEUS FUR simulation has already allowed us to infer a number of properties of the distribution of matter in the Universe which are relevant for future observational studies on Dark Energy. 

\begin{figure}[!h]
	\centering
	\includegraphics[scale=0.35]{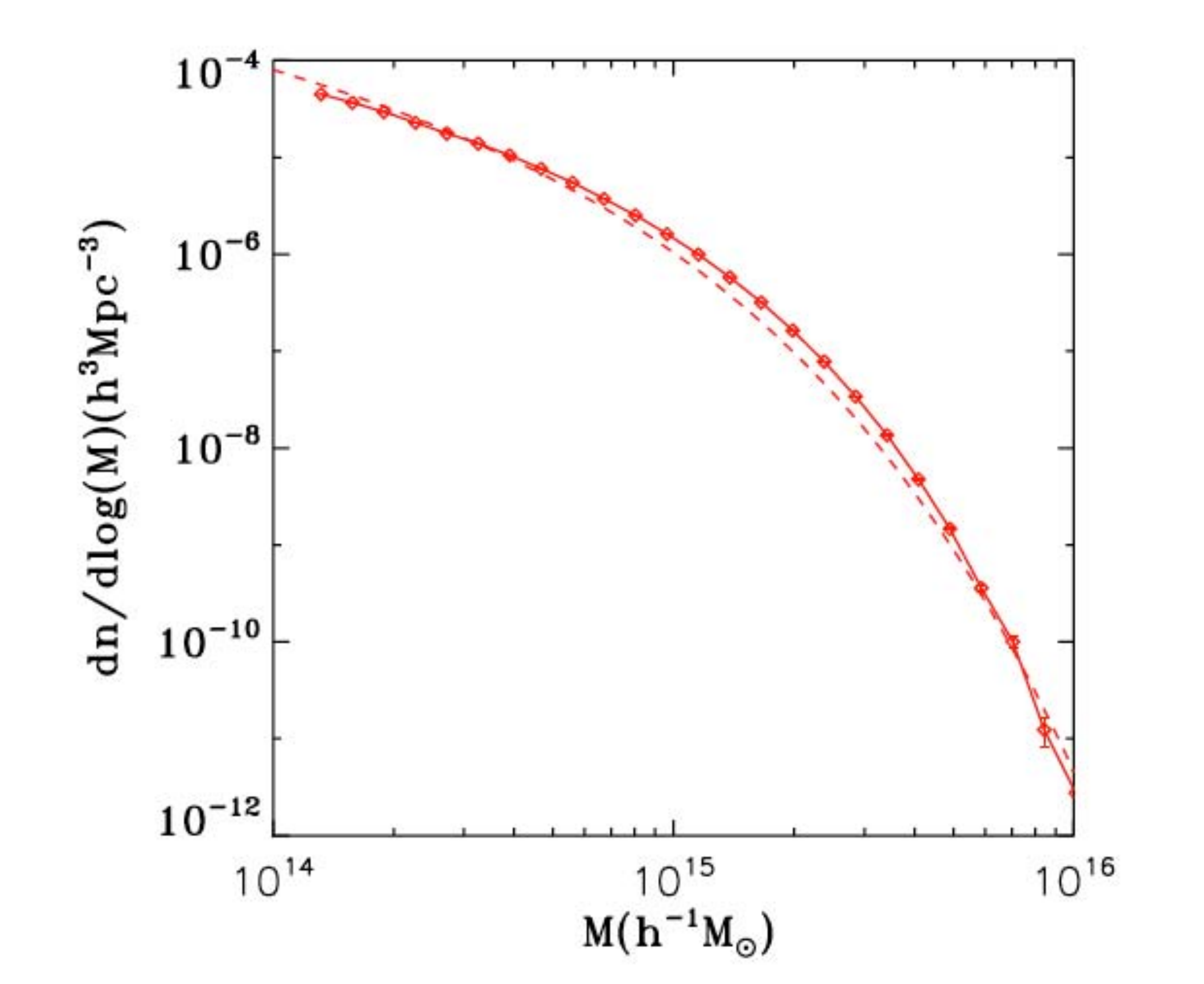}
	\caption{Mass function of massive halos detected in the simulation. The dashed line corresponds to Jenkins et al. \ref{jenkins} fitting function.}
	\label{Massfunction}
\end{figure}

First, we have computed the total mass function of DM halos (see figure \ref{Massfunction}). Especially at large masses, the analysis confirms the presence of deviations from universality \cite{Courtin2011}. This widely used approximation assumes that the mass function can be expressed with no explicit dependence on the properties of Dark Energy or in time. In contrast, our results suggests that measurements of the halo mass function today can provide additional constraints on dark energy since it carries a fossil record of the past cosmic evolution. 

We have detected more than $144$ millions DM halos with mass larger than $10^{14}$ M$_\odot$. The first halo of at least $10^{14}$ M$_\odot$ has formed
when the universe was only 2 billion years old. The most massive halo in the observable universe today has a mass of $15$ quadrillion M$_\odot$.

\begin{figure}[!h]
	\centering
	\includegraphics[scale=0.25]{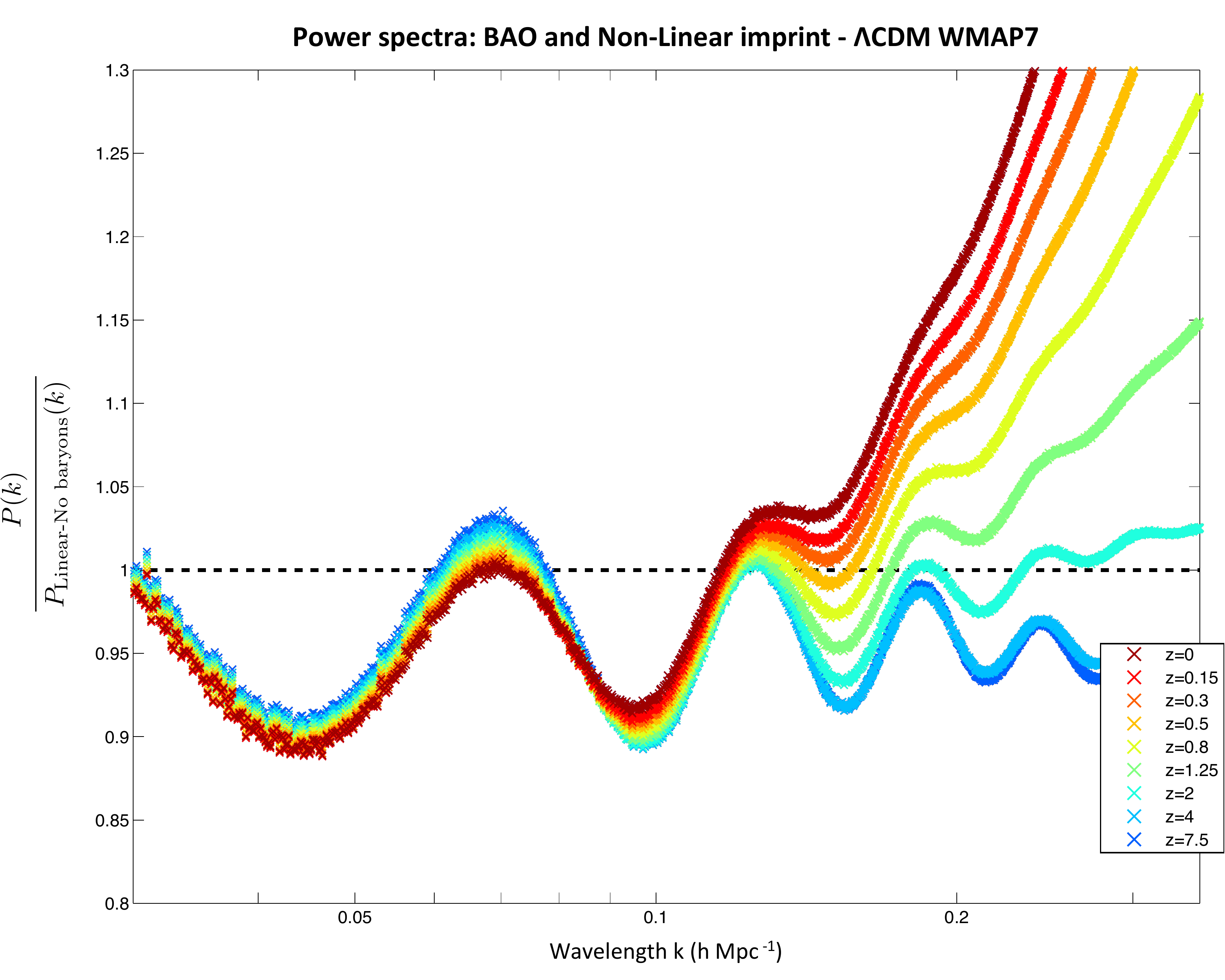}
	\caption{Evolution of the imprint of the ``baryon acoustic oscillations'' in the density matter power spectrum fluctuations. The position and height of the peaks measured in the simulation with unprecedented precision allows to constrain the cosmological model, especially the amount and the nature of Dark Energy. At small scales (large k-values) we observe the deformation due to non-linear dynamics.}
	\label{BAO}
\end{figure}

The data generated by the run have also allowed us to evaluate the spatial distribution of dark matter density fluctuations in the universe. In particular, 
the analysis of the matter power spectrum reveals with unprecedented accuracy the imprint of the primordial plasma's acoustic oscillations on the Dark Matter distribution (Baryon Acoustic Oscillations). We can clearly in Figure \ref{BAO} the deformations of the oscillations by the non-linear evolution at small scales and at different times.

\begin{figure}[h]
	\centering
	\includegraphics[scale=0.18]{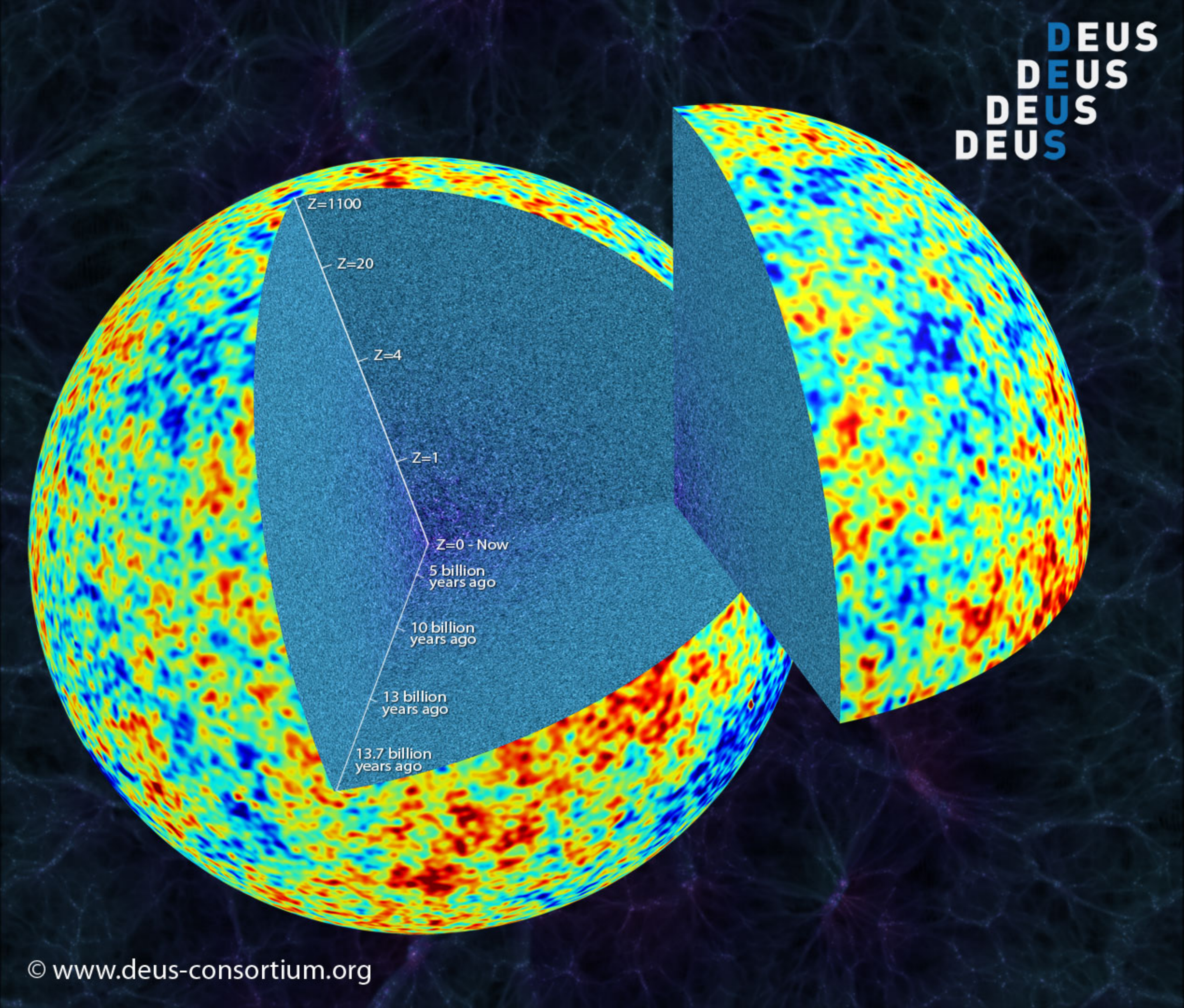}
	\caption{First picture of the 3D dark matter distribution in redshift space resulting from the evolution of the density matter fluctuations observed by WMAP satellite in the $\Lambda CDM$ Concordance Cosmological Model with Cosmological Constant.}
	\label{DEUSTranche}
\end{figure}

\begin{figure}[!b]
	\centering
	\includegraphics[scale=0.45]{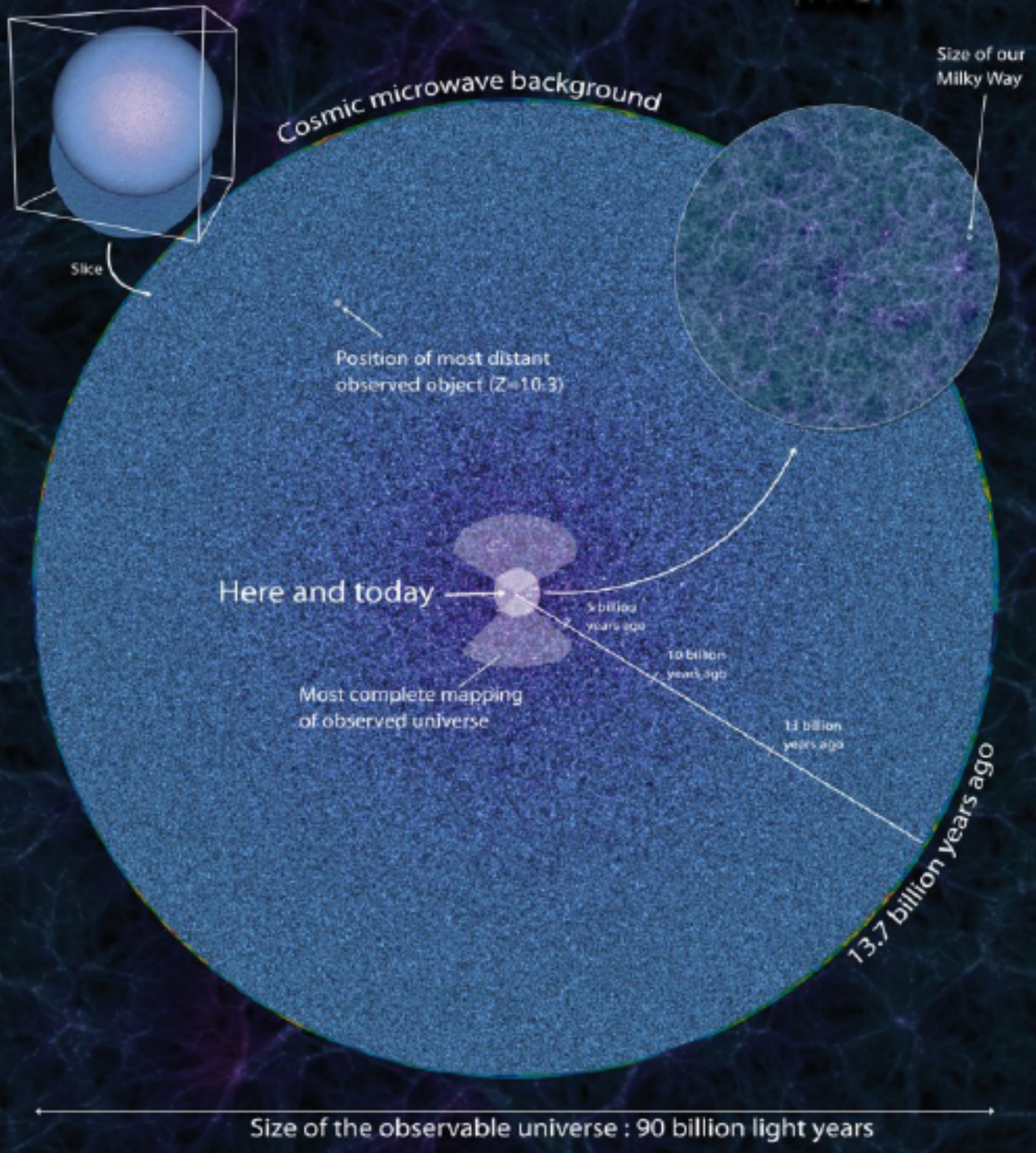}
	\caption{Volume of the universe accessible through simulation, i.e. the full observable universe. Also shown is the comparison with the volume extent of the universe currently observed. At the edge of the disk extracted from the full celestial sphere, we find the Cosmic Microwave Background.}
	\label{lightcone}
\end{figure}

For the first time, using the capabilities of the CURIE supercomuter, the simulation DEUS FUR allows us to reproduce the full sky Dark Matter distribution in 3D redshift space all the way to the CMB (see Figure \ref{DEUSTranche}). In Figure \ref{lightcone}) it is shown a 2D spatial section of the full sky ``light-cone''.

The data generated by the run will provide an exceptional support to future surveys of the large scales structures which will probe the distribution
of matter over larger cosmic volumes.

\section*{Acknowledgment}
We are very thankful to R. Teyssier (CEA, IRFU for original RAMSES code), P. Wautelet (IDRIS for some optimization of original RAMSES Code), S. Prunet \& C. Pichon (IAP for MPGRAFIC), F. Roy (LUTh for his developments on FOF), J. Pasdeloup (LUTh for Visualization development), B. Frog\'e, F. Diakhate and M. Hautreux (CEA, TGCC for Technical Support on CURIE supercomputer).



%


\end{document}